\documentclass{article}
\usepackage{spconf,amsmath,graphicx}

\usepackage{enumitem}
\setlist{nosep, leftmargin=14pt}

\usepackage{graphicx,import}
\usepackage{cite}
\usepackage{algorithmic}
\usepackage{textcomp}
\usepackage{xcolor}
\usepackage{booktabs}
\usepackage{blindtext}
\def\BibTeX{{\rm B\kern-.05em{\sc i\kern-.025em b}\kern-.08em
    T\kern-.1667em\lower.7ex\hbox{E}\kern-.125emX}}

\makeatletter
\def\bstctlcite{\@ifnextchar[{\@bstctlcite}{\@bstctlcite[@auxout]}}
\def\@bstctlcite[#1]#2{\@bsphack
  \@for\@citeb:=#2\do{%
    \edef\@citeb{\expandafter\@firstofone\@citeb}%
    \if@filesw\immediate\write\csname #1\endcsname{\string\citation{\@citeb}}\fi}%
  \@esphack}
\makeatother

\begin{document}
\bstctlcite{IEEEexample:BSTcontrol}

\title{Anatomical Conditioning for Contrastive Unpaired Image-to-Image Translation of Optical Coherence Tomography Images}

 \name{Marc S. Seibel$^{\star}$ \qquad Hristina Uzunova$^{\dagger}$ \qquad Timo Kepp$^{\dagger}$ \qquad Heinz Handels$^{\star\dagger}$}

 \address{$^{\star}$ Institute of Medical Informatics, University of Lübeck, Germany  \\
     $^{\dagger}$German Research Center for Artificial Intelligence (DFKI), Lübeck, Germany}

\maketitle

\begin{abstract}
For a unified analysis of medical images from different modalities, data harmonization using image-to-image (I2I) translation is desired. We study this problem employing an optical coherence tomography (OCT) data set of Spectralis-OCT and Home-OCT images. I2I translation is challenging because the images are unpaired, and a bijective mapping does not exist due to the information discrepancy between both domains. This problem has been addressed by the Contrastive Learning for Unpaired I2I Translation (CUT) approach, but it reduces semantic consistency. To restore the semantic consistency, we support the style decoder using an additional segmentation decoder. Our approach increases the similarity between the style-translated images and the target distribution. Importantly, we improve the segmentation of biomarkers in Home-OCT images in an unsupervised domain adaptation scenario. Our data harmonization approach provides potential for the monitoring of diseases, e.g., age related macular disease, using different OCT devices.

\end{abstract}


%\begin{IEEEkeywords}
%component, formatting, style, styling, insert
%\end{IEEEkeywords}

\section{Introduction}
Image-to-Image  style translation (I2I) using Generative Adversarial Networks (GANs) has found wide applicability for building robust models \cite{uzunova2020generation,shinMedicalImageSynthesis2018}. I2I has been used to reduce the domain shift for downstream applications such as image segmentation \cite{platscherImageTranslationMedical2022} or image registration \cite{chenUnsupervisedMultiModalMedical2022}. 
By reducing domain shifts, I2I is a promising tool for transfer learning \cite{pangImagetoImageTranslationMethods2022} which is a prevalent application in deep learning for medical imaging where data sets are often small and come with biases \cite{willeminkPreparingMedicalImaging2020}.   
However, in cases where no paired image data exists, translating images is challenging \cite{gutierrezLesionpreservingUnpairedImagetoimage2023}. An example of such a scenario is the analysis of optical coherence tomography (OCT) images. Patients might undergo OCT monitoring due to diseases such as age related macular degeneration (AMD). AMD is currently diagnosed in clinical environments. To examine the state of the disease on a daily basis, Home-OCT has been invented \cite{sudkampInvivoRetinalImaging2016}. To automatically assess the progress of the disease using the clinical scans and the Home-OCT images, it is desirable that the images from both devices are comparable. However, harmonizing the style of these images is difficult since different acquisition techniques give rise to distinct representations of noise and movement related artifacts \cite{bogunovicRETOUCHRetinalOCT2019,kochNoiseTransferUnsupervised2022a}. Consequently, the information content in OCT images differs depending on the device. 
%This is particularly infeasible with medical image data due to its sensitive nature, since a hallucinated structure might be interpreted as pathological in an originally healthy control. 
Nonetheless, it can intuitively be assumed that a bidirectional mapping for unpaired data still exists, i.e., a function mapping a source image to a given target image unambiguously such that the inverse of the function can be determined. This assumption is the basis of models like CycleGAN \cite{zhuUnpairedImagetoImageTranslation2017}, where a cycle-consistency loss ensures that the target image can be translated to the source image and back to its original and vice versa. Although such models have proven to be successful for various medical image applications \cite{armaniousUnsupervisedMedicalImage2019}, the bijective assumption might not always hold \cite{parkContrastiveLearningUnpaired2020a}. In fact, for OCT, manually designed methods provide a style transfer that is more useful for downstream semantic segmentation than CycleGAN\cite{kochNoiseTransferUnsupervised2022a}. We attribute this to the large portions of missing information, which makes it infeasible to find bijective functions. \par
This problem is addressed by the Contrastive Learning for Unpaired Image-to-Image translation (CUT) approach \cite{parkContrastiveLearningUnpaired2020a}, where the authors propose a contrastive learning strategy as an alternative to cycle-consistency. This way they maintain correspondence in content but not appearance by maximizing the mutual information between corresponding input and output patches.
One shortcoming of the CUT approach compared to CycleGAN is the susceptibility to data set imbalances \cite{parkContrastiveLearningUnpaired2020a}. Data set imbalances are a reason for structure hallucination, when distribution matching approaches are employed for I2I translation \cite{cohenDistributionMatchingLosses2018}. 
%In the original CUT paper, the authors showed this effect with a data set of horses and zebras where the size style transfered images . Data set imbalances are a prominent problem in medical images, since some classes (healthy patients vs. patients with a particular disease) might naturally appear more often, some acquisition methods might be more favored due to their efficiency or different acquisition devices consider different fields of view.
% Hierhin related works!

To cope with the above-mentioned problems, we introduce anatomically conditioned contrastive unpaired image-to-image translation. Our method extends the CUT approach by introducing additional anatomical conditioning, which is intended to suppress the hallucination of structures. The anatomical conditioning is implemented as an additional segmentation decoder, that shares features on multiple resolutions with the style decoder (Fig.~\ref{fig:network}). 
%Here, the properties of this approach are studied on retinal OCT images, with the specific use-case being the domain translation between Spectralis-OCT and Home-OCTs.
 \par

%Approaches for I2I translation can be categorized acc
%One limiting factor of GAN-based approaches remains their proneness to hallucinate structures. 

%is one way to reduce the domain gap

%Cycle GANs 
%However, the cycle consistency assumption does not hold true if In CUT, the authors describe one critical 

%Here, we introduce anatomical conditioned contrastive unpaired image-to-image translation to address the problem of hallucination.

\section{Method}
\begin{figure}[!tbp]
\centering
\includegraphics[width=3in]{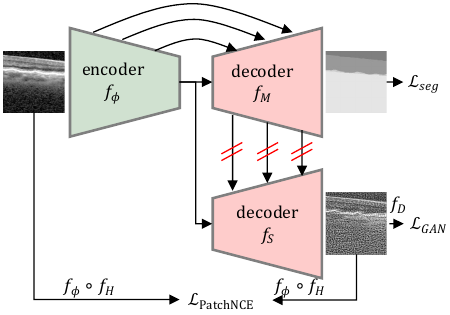}
\caption{Anatomical conditioning for style transfer. The segmentation decoder $f_M$ provides shape information to the style decoder $f_S$. The red lines denote that optimizing the style transfer loss does not update the segmentation decoder (no backpropagation).}
\label{fig:network}
\end{figure}
Our method, anatomically conditioned CUT (ACCUT), is schematically presented in Fig.~\ref{fig:network}. As in CUT, the model consists of a generator with an encoder $f_\phi$ and a style decoder $f_S$, a discriminator $f_D$, and a two-layer MLP $f_H$ for producing features which are used for the contrastive learning loss. The discriminator learns to differentiate between fake and real target images and is used to update the generator in an adversarial fashion. We propose to use a semantic segmentation decoder $f_M$ to guide the style decoder $f_S$ with information about the shape and topology of the objects based on a segmentation mask. Since $f_\phi$ is shared for the semantic segmentation and the style transfer task, it learns features which are relevant for both. To combine the features learned by both decoders, the multi-resolution features from $f_M$ are concatenated to the features of $f_S$. The aim for $f_M$ is to only represent shape information which is invariant to the appearance of the input image. Therefore, we exclude $f_M$ from backpropagation when optimizing the style transfer loss.  

Our loss function extends the CUT loss with two terms for the semantic segmentation. Categorical cross entropy is chosen as segmentation loss
\begin{equation}
    \mathcal{L}_{seg}(f_\phi, f_M, \hat y, {y}) = -\sum_{i=1}^{C} y_i \cdot \log(\hat{y}_i),
    \label{eq:seg_loss}
\end{equation}
where $y$ denotes the ground truth masks and $\hat{y}$ the predicted segmentations. The final objective function is the sum of the conventional CUT loss and the segmentation losses
\begin{equation}
\begin{split}
    \mathcal{L} &=   \mathcal{L}_{\text{GAN}}\left(f_\phi,\,f_S,\, f_D,\, X_t, X_s \right) \\& + \lambda_{X_s}\mathcal{L}_{\text{PatchNCE}}\left(f_\phi,\,f_S,\,f_H,\,X_s\right)\\&+
    \lambda_{X_t}\mathcal{L}_{\text{PatchNCE}}\left(f_\phi,\,f_S,\,f_H,\,X_t\right)\\& + \lambda_{s}\mathcal{L}_{\text{seg}}\left(f_\phi,\,f_M,\,\hat y_s,\,{y}_s\right) + \lambda_{t}\mathcal{L}_{\text{seg}}\left(f_\phi,\,f_M,\,\hat y_t,{y}_t\right).
    \end{split}
    \label{eq:total}
\end{equation}
The losses $\mathcal{L}_{\text{GAN}}$ and $\mathcal{L}_{\text{PatchNCE}}$ correspond to the adversarial loss and to the patch-based contrastive loss in \cite{parkContrastiveLearningUnpaired2020a} where $X_t$ and $X_s$ denote the target and source images, respectively. In the segmentation loss terms, $y_s$ and $y_t$ denote the ground truth masks and $\hat{y_t}$ and $\hat{y_s}$ the predicted segmentations for target and source images.
Throughout our experiments, we set the weights  $\lambda_{X_t}=\lambda_{X_s}=1.0$. Since the value of $\lambda_s$ and $\lambda_t$ corresponds to the usage of semantic information, we define four operating modes i.e. CUT ($\lambda_s=\lambda_t=0.0$), ACCUT$_s$ ($\lambda_s=1.0,\lambda_t=0.0$), ACCUT$_t$ ($\lambda_s=0.0,\lambda_t=1.0$), and ACCUT$_{s,t}$ ($\lambda_s=\lambda_t=0.5$). Notably, we use the same network architecture and weights to calculate the segmentation for the source and target images. 

For the model architecture, the ResNet-based network described in \cite{parkContrastiveLearningUnpaired2020a} is used. However, we modify the bottleneck so that the 6-block ResNet is split into four blocks for the encoder and two blocks were given to the style and semantic segmentation decoder each. Our implementation can be found on GitHub~\footnote{https://github.com/msseibel/ACCUT}.

\section{Experiments}

\begin{figure*}[!t]
\centering
\includegraphics[width=\linewidth]{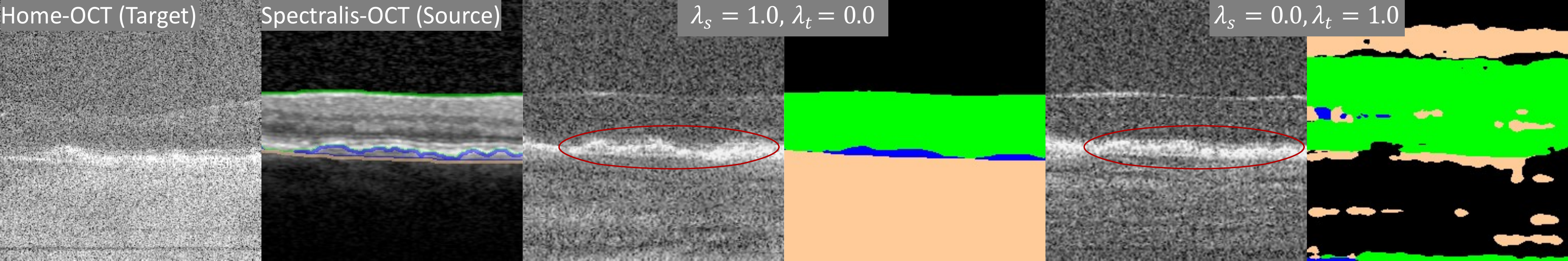}
\caption{Style transfer and simultaneous segmentation with ACCUT. From left to right, we show a Home-OCT image from the target domain, a Spectralis image from the source domain with its ground truth segmentation, the Spectralis image translated to the target domain and its corresponding segmentation using ACCUT$_{s}$ and ACCUT$_{t}$, respectively. Training the segmentation decoder with only target domain supervision results in bad style transfer for the PED (red circle).}
\label{fig:ds_errs}
\end{figure*}
For evaluating the ACCUT approach, we use a data set consisting of Spectralis-OCT (Heidelberg Engineering GmbH) and Home-OCT (Visotec GmbH) images \cite{burchard}. The data set features 38 subjects, which were each examined with both devices. The included subjects were diagnosed with eye diseases, of which neovascular AMD was most common. Subretinal fluids (SRF) and pigment epithelial detachment (PED) were annotated by a clinical expert as relevant AMD biomarkers \cite{burchard}. Additionally, the retina was also annotated, dividing the OCT image into two parts. We automatically annotate the area above and below the retina as vitreous body and choroid, respectively. The Spectralis-OCT images were resampled to the same resolution as the Home-OCT images. Examples of a Home-OCT and Spectralis are given in Fig.~\ref{fig:ds_errs}.

The network for ACCUT was trained with the Spectralis-OCT as source and the Home-OCT as target data on the whole data set. The model was evaluated after 250 training epochs, which is an empirically found value. For evaluation, we seek to answer four questions:  
(Q1) Does ACCUT improve downstream segmentation results in an unsupervised domain adaptation (UDA) setting where only the source images were translated to the target domain?  
(Q2) Are the domain-translated images more similar to the target domain when using ACCUT compared to the conventional CUT?
(Q3) Should semantic conditioning be applied to the source, target or both  domains?
(Q4) Is the style decoder using the information from the mask decoder, respectively,  does the anatomical conditioning influence the results?

\subsection{ACCUT for Domain Adaptation}

To answer, whether the ACCUT style transfer improves semantic segmentation as a downstream task (Q1), we trained the EfficientNet-B2 \cite{tanEfficientNetRethinkingModel2019} on the Spectralis-OCTs and their domain-translated versions and tested on real Home-OCT images. We evaluated six versions of training data (1) no style translation (Spectralis-OCTs); (2) style translation using the CycleGAN; (3) style translation using CUT; (4) ACCUT$_s$ with $\lambda_s=1.0$ and $\lambda_t=0$, hence, only segmentation of the source images; (5) ACCUT$_t$ with $\lambda_s=0$ and $\lambda_t=1.0$, hence, only segmentation of the target images; (6) ACCUT$_{s,t}$ with $\lambda_s=0.5$ and $\lambda_t=0.5$, including segmentation of source and target data. Note that (5) and (6) are generally not relevant for the unsupervised domain adaptation scenario, since no labels for the target domain exist by definition. Nonetheless,  we want to evaluate whether complete semantic supervision creates images which are more useful for training the domain adaptation model.

The network was trained with 5-fold cross-validation. Models were trained for 15 epochs and based on the source validation data, the best model is selected for testing on the target test data. For all six experiments, slight data augmentation was used: flipping around the vertical axis, randomly changing the resolution, random crops of size (256, 480), random gamma transformation and histogram shifting. As the loss function, categorical cross entropy was used and weights are updated with the Adam optimizer.\par
The results of this experiment are shown in Tab.~\ref{tab:my-table}. The segmentation results are evaluated in terms of Dice coefficient per structure and mean Dice coefficient over all folds and subjects. Regarding Q3, the best average segmentation is achieved with the domain translation using ACCUT$_s$ and ACCUT$_{s,t}$ achieving 60~\% segmentation accuracy compared to 40\% when no domain translation is applied and 57~\% using the conventional CUT approach. Using only anatomical restrains for the target image with ACCUT$_t$, however, does not lead to segmentation improvement. This is due to the fact that the segmentation accuracy of ACCUT$_t$ is poor (Fig.~\ref{fig:ds_errs}). This underlines the necessity for anatomical conditioning on the source image domain. It should be noted that PEDs and SRFs are small structures, making their segmentation a challenging task. 
\setlength{\tabcolsep}{5pt}
\begin{table}[!thb]
\centering
\begin{tabular}{llllll|ll}
\toprule
           %& \multicolumn{6}{c}{Dice [\%]}                     
           method & V. & C. & R. & SRF & PED & mDice $\uparrow$ & FID $\downarrow$ \\ \midrule
Source   & 75        & 72     & 46    & 1 & 8 & 40 $\pm$ 8 & 248\\
CycleGAN   & 83        & 82     & 47    & 1 & 7 & 44 $\pm$ 8 & 99\\
CUT           & 91        & \textbf{93}     & 78    & 10 & 14 & 57 $\pm$ 3 & 121\\
ACCUT$_{s,t}$    & \textbf{92}        & \textbf{93}     & \textbf{86}    & 7 & \textbf{20} & \textbf{60} $\pm$ 4 & 105\\
ACCUT$_{s}$    & \textbf{92}        & 92     & 81    & \textbf{18} & 18 & \textbf{60} $\pm$ 4 & \textbf{89}\\
ACCUT$_{t}$   & 90        & 92     & 73    & 2 & 7 & 53 $\pm$ 3 & 120\\ \bottomrule
\end{tabular}
\caption{Results for the segmentation UDA downstream task (Q1) and image similarity (Q2) using the different domain translation methods. Segmentation quality is given as Dice score [\%] per structure: vitreous body (V.), choroid (C.), retina (R.), PED and SRF. Also, the mean Dice and its standard deviation over all folds are given (mDice). Image similarity is given by the Fréchet Inception Distance (FID). }
\label{tab:my-table}
\end{table}

\subsection{Ablation Study}
\begin{figure*}[!h]
\centering
\includegraphics[width=\linewidth]{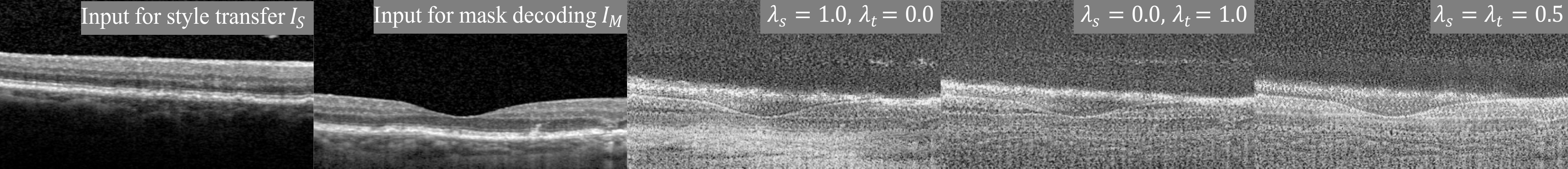}
\caption{Ablation study. A non-matching input is used for the mask decoder, and its features are concatenated to the style decoder. The translated images contain the images from both input images, which shows that the style decoder uses the information from the mask decoder.}
\label{fig:ablation}
\end{figure*}
This experiment aims to analyze the role that the anatomical conditioning plays in the style transfer, i.e. whether the concatenated features in the style decoder simply get ignored or influence the style transfer of the network (Q4). The experiment is designed in the form of an ablation study where two different images $I_S$ and $I_M$ are fed to the network. Then, the encoded features $f_\phi(I_S)$ are only used as inputs for the style decoder, whereas the features $f_\phi(I_M)$ are only passed to the segmentation decoder. The concatenation of the multi-resolution decoder feature maps is subsequently carried out as usual. In this way, non-corresponding masks are used and the effect of the anatomical conditioning can be visualized (Fig.~\ref{fig:ablation}). It becomes apparent that the mask decoding plays a crucial role in the final style-transferred image. Using different $\lambda_s$ and $\lambda_t$ results in different degrees of anatomical constraining, however, the tendency of maintaining the shape of the image $I_M$ is visible in all examples. The style-transferred images roughly follow the topology of the anatomical layers and the overall retinal position of $I_M$. Hence, this study not only shows that the proposed anatomical conditioning plays an important role in the final generated image. It also prevents the problem of affine deformation imbalance of the different domains (e.g., one of the domains contains generally larger structures) which is a major concern for the conventional CUT method.
%We evaluated whether the concatenation of the segmentation features effected the style decoding by feeding two different images to the decoder. The latent features of the first images were given to the style decoder, while the latent features of the second image were only passed to the segmentation decoder. The decoded segmentation features from the second image were then concatenated with the multi-resolution feature maps of the first image. We see in~\ref{fig:ablation} that the decoded image contains structures from both images.
%\begin{table}[]
%\centering
%\begin{tabular}{llllllll|ll}
%\toprule
           %& \multicolumn{6}{c}{Dice [\%]}                       \\
%           method &$\lambda_s$&$\lambda_t$& G. & C. & R. & PED & \multicolumn{1}{l|}{SRF} & \multicolumn{1}{l}{mDice} & FID \\ \midrule
%source only& -  & -  & 75        & 72     & 46    & 1 & 8 & 40 & 248\\
%CUT        & 0.0  & 0.0 & 91        & 93     & 78    & 10 & 14 & 57 & 121\\
%ACCUT $\vert$ S,T & 0.5&0.5    & 92        & 93     & 86    & 7 & 20 & 60 & 105\\
%ACCUT $\vert$ S & 1.0&0.0    & 92        & 92     & 81    & 18 & 18 & 60 & 89\\
%ACCUT $\vert$ T & 0.0&1.0    & 90        & 92     & 73    & 2 & 7 & 53 & 120\\ \bottomrule
%\end{tabular}
%\caption{}
%\label{tab:my-table}
%\end{table}

% Why does CUT| S, T not work and why not CUT|T

\subsection{Image Similarity}
\begin{figure}[!b]\centering
\def\svgwidth{2.5in} % or another length
%% Creator: Inkscape 1.3 (0e150ed6c4, 2023-07-21), www.inkscape.org
%% PDF/EPS/PS + LaTeX output extension by Johan Engelen, 2010
%% Accompanies image file 'qualitative_tex.pdf' (pdf, eps, ps)
%%
%% To include the image in your LaTeX document, write
%%   \input{<filename>.pdf_tex}
%%  instead of
%%   \includegraphics{<filename>.pdf}
%% To scale the image, write
%%   \def\svgwidth{<desired width>}
%%   \input{<filename>.pdf_tex}
%%  instead of
%%   \includegraphics[width=<desired width>]{<filename>.pdf}
%%
%% Images with a different path to the parent latex file can
%% be accessed with the `import' package (which may need to be
%% installed) using
%%   \usepackage{import}
%% in the preamble, and then including the image with
%%   \import{<path to file>}{<filename>.pdf_tex}
%% Alternatively, one can specify
%%   \graphicspath{{<path to file>/}}
%% 
%% For more information, please see info/svg-inkscape on CTAN:
%%   http://tug.ctan.org/tex-archive/info/svg-inkscape
%%
\begingroup%
  \makeatletter%
  \providecommand\color[2][]{%
    \errmessage{(Inkscape) Color is used for the text in Inkscape, but the package 'color.sty' is not loaded}%
    \renewcommand\color[2][]{}%
  }%
  \providecommand\transparent[1]{%
    \errmessage{(Inkscape) Transparency is used (non-zero) for the text in Inkscape, but the package 'transparent.sty' is not loaded}%
    \renewcommand\transparent[1]{}%
  }%
  \providecommand\rotatebox[2]{#2}%
  \newcommand*\fsize{\dimexpr\f@size pt\relax}%
  \newcommand*\lineheight[1]{\fontsize{\fsize}{#1\fsize}\selectfont}%
  \ifx\svgwidth\undefined%
    \setlength{\unitlength}{375.00005767bp}%
    \ifx\svgscale\undefined%
      \relax%
    \else%
      \setlength{\unitlength}{\unitlength * \real{\svgscale}}%
    \fi%
  \else%
    \setlength{\unitlength}{\svgwidth}%
  \fi%
  \global\let\svgwidth\undefined%
  \global\let\svgscale\undefined%
  \makeatother%
  \begin{picture}(1,1)%
    \lineheight{1}%
    \setlength\tabcolsep{0pt}%
    \put(0,0){\includegraphics[width=\unitlength,page=1]{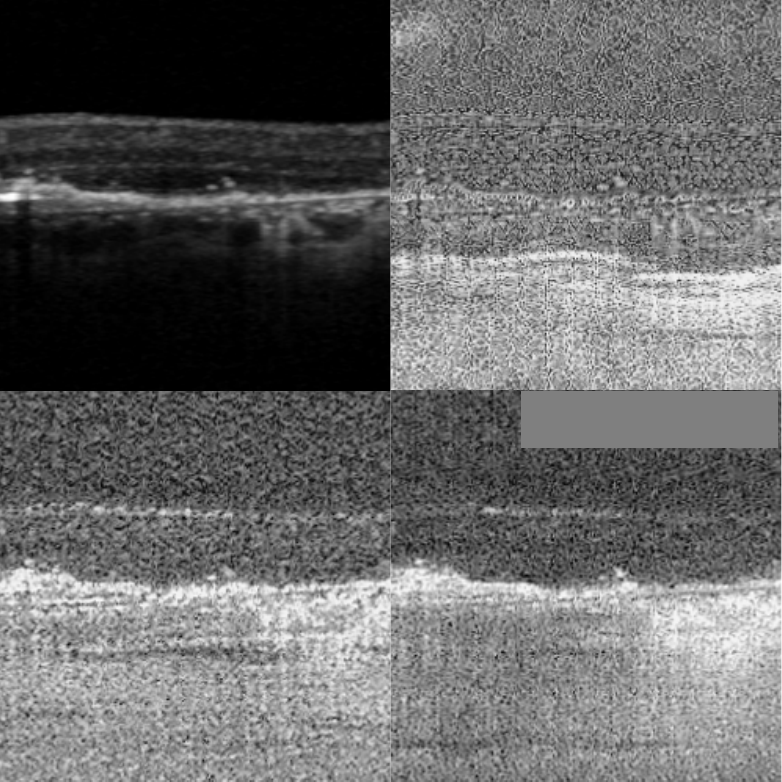}}%
    \put(0.73108976,0.45081929){\color[rgb]{1,1,1}\makebox(0,0)[lt]{\lineheight{1.25}\smash{\begin{tabular}[t]{l}ACCUT$_s$\end{tabular}}}}%
    \put(0,0){\includegraphics[width=\unitlength,page=2]{qualitative_tex.pdf}}%
    \put(0.28217168,0.44559012){\color[rgb]{1,1,1}\makebox(0,0)[lt]{\lineheight{1.25}\smash{\begin{tabular}[t]{l}CUT\end{tabular}}}}%
    \put(0,0){\includegraphics[width=\unitlength,page=3]{qualitative_tex.pdf}}%
    \put(0.71524453,0.9504729){\color[rgb]{1,1,1}\makebox(0,0)[lt]{\lineheight{1.25}\smash{\begin{tabular}[t]{l}CycleGAN\end{tabular}}}}%
    \put(0,0){\includegraphics[width=\unitlength,page=4]{qualitative_tex.pdf}}%
    \put(0.22991908,0.95042081){\color[rgb]{1,1,1}\makebox(0,0)[lt]{\lineheight{1.25}\smash{\begin{tabular}[t]{l}Spectralis\end{tabular}}}}%
  \end{picture}%
\endgroup%

\caption{Qualitative comparison of the image translation methods.}
\label{fig:images_qual}
\end{figure}
To assess the similarity of the real target images and the generated ones (Q2), the commonly used Fréchet Inception Distance (FID) is utilized \cite{heuselGANsTrainedTwo2017}. The smallest values are achieved by the ACCUT$_s$ approach, implying the best similarity between the real and the generated images (Tab.~\ref{tab:my-table}). With much higher FID-values, ACCUT$_t$ and CUT seem to yield the worst domain translation results, which also corresponds to the worst segmentation accuracy achieved with the approaches. A qualitative comparison is given in \ref{fig:images_qual}. One can see that the CycleGAN fails to correctly find the transition from the retina to the choroid. Additionally, in ACCUT$_s$ the texture of the retina is better differentiable from the vitreous body above the retina compared to CUT.

\section{Conclusion}
% Why does CUT| S, T not work and why not CUT|T
In this work, we present an extended CUT framework which additionally considers semantic information by employing a segmentation decoder next to the style decoder. Our experiments demonstrate that this additional guidance improves the quality of domain translated images by addressing data set imbalances. The subsequent ablation study shows that the style decoder incorporates the anatomical information. Furthermore, the improved segmentation quality in the downstream UDA experiment on the Visotec target data set demonstrated that the Spectralis images were translated more succesfully to the Visotec domain when using ACCUT$_s$ compared to CUT. In particular, the clinical relevant biomarkers SRF and PED were better segmented using ACCUT$_s$. The ACCUT$_t$ model lead to a worse UDA segmentation compared to CUT which could be explained by the bad segmentation quality for the source images which then resulted in a diminished style transfer quality and consequently lead to worse training images for the UDA model. However, the  impact of source and target segmentations on ACCUT needs to be studied, since it is likely that the optimal choice of loss weights depends on the choice of the source and target data set.

Future work will concentrate on the introduction of further data sets from the medical image domain. Also, architectural choices will be investigated more  profoundly. For example, the simple concatenation of the features of the style and segmentation decoders can be replaced by more elaborate mechanisms such as attention. 
Even though there is a significant potential for further developments, ACCUT provides a basic methodology to address the class imbalance and hallucination problem in image-to-image translation.
%Additionally, integrating ACCUT into an end-to-end learning pipeline for unsupervised domain adaptation is planned, as in Cycada \cite{hoffmanCyCADACycleConsistentAdversarial2018a}. 

% Limitation: 
% 1. More data sets needed
% 2. The network architecture is not fully optimized: 
% - Attention

% integrating ACCUT directly into UDA as in the cycada approach.

\section*{Acknowledgment}
M. Seibel was supported through the GAIA-X-MED project funded by the federal state Schleswig-Holstein. Finally, we thank our colleagues at the DFKI, the IMI-UzL, and Visotec GmbH for their input throughout the duration of this work.

\section*{Compliance with Ethical Standards}
Recording of the data used in this study, was approved by Ethics Committee, University of Kiel (reference number: A139/17). Participants gave informed consent to participate in the study before taking part.
%\section*{References}

% \begin{thebibliography}{00}
% \bibitem{b1} G. Eason, B. Noble, and I. N. Sneddon, ``On certain integrals of Lipschitz-Hankel type involving products of Bessel functions,'' Phil. Trans. Roy. Soc. London, vol. A247, pp. 529--551, April 1955.
% \bibitem{b2} J. Clerk Maxwell, A Treatise on Electricity and Magnetism, 3rd ed., vol. 2. Oxford: Clarendon, 1892, pp.68--73.
% \bibitem{b3} I. S. Jacobs and C. P. Bean, ``Fine particles, thin films and exchange anisotropy,'' in Magnetism, vol. III, G. T. Rado and H. Suhl, Eds. New York: Academic, 1963, pp. 271--350.
% \bibitem{b4} K. Elissa, ``Title of paper if known,'' unpublished.
% \bibitem{b5} R. Nicole, ``Title of paper with only first word capitalized,'' J. Name Stand. Abbrev., in press.
% \bibitem{b6} Y. Yorozu, M. Hirano, K. Oka, and Y. Tagawa, ``Electron spectroscopy studies on magneto-optical media and plastic substrate interface,'' IEEE Transl. J. Magn. Japan, vol. 2, pp. 740--741, August 1987 [Digests 9th Annual Conf. Magnetics Japan, p. 301, 1982].
% \bibitem{b7} M. Young, The Technical Writer's Handbook. Mill Valley, CA: University Science, 1989.
% \end{thebibliography}
%\printbibliography
\bibliographystyle{IEEEtran}
\bibliography{IEEEabrv, literature, vonderburchard, settings}{} % , spie_seibel, vonderburchard

\end{document}